\documentclass[final,referee]{paper}

\usepackage{todonotes}
\usepackage[mathlines]{lineno}
\newcommand*\patchAmsMathEnvironmentForLineno[1]{\expandafter\let\csname old#1\expandafter\endcsname\csname #1\endcsname
  \expandafter\let\csname oldend#1\expandafter\endcsname\csname end#1\endcsname
  \renewenvironment{#1}{\linenomath\csname old#1\endcsname}{\csname oldend#1\endcsname\endlinenomath}}\newcommand*\patchBothAmsMathEnvironmentsForLineno[1]{\patchAmsMathEnvironmentForLineno{#1}\patchAmsMathEnvironmentForLineno{#1*}}\AtBeginDocument{\patchBothAmsMathEnvironmentsForLineno{equation}\patchBothAmsMathEnvironmentsForLineno{align}\patchBothAmsMathEnvironmentsForLineno{flalign}\patchBothAmsMathEnvironmentsForLineno{alignat}\patchBothAmsMathEnvironmentsForLineno{gather}\patchBothAmsMathEnvironmentsForLineno{multline}}
\usepackage{empheq}

\usepackage{amsmath}
\usepackage{mathtools}
\usepackage{amssymb} \usepackage{mathrsfs} \usepackage{IEEEtrantools} \usepackage{fp}   

\usepackage[subnum]{cases} \usepackage{cancel}

\usepackage{xstring}

\def\u2#1{\underline{\underline{#1}}}  \newcommand{\bs}[1]{\boldsymbol{#1}}

 \newtheorem{remark}{Remark} 
\usepackage[shortlabels]{enumitem} \usepackage{pifont}     

\usepackage{parskip} \usepackage[normalem]{ulem} \usepackage{accents}
\usepackage{csquotes} \def\colorize<#1>{\temporal<#1>{\color{black!30}}{\color{red}}{\color{black}}}
\usepackage{varwidth,lipsum} \usepackage{tabto} \usepackage{tabstackengine}

\usepackage{fullpage}

\usepackage[percent]{overpic} \usepackage[rightcaption]{sidecap}
\usepackage{caption}
\usepackage{subcaption}
\usepackage{graphicx} 

\usepackage{epstopdf} \usepackage{latexsym}\usepackage{keyval}\usepackage{ifthen}\usepackage{moreverb}\usepackage[subfolder]{gnuplottex}

\usepackage{tikz}
\usepackage{tikz, calc, overpic, fp}

\usetikzlibrary{decorations.markings}

\graphicspath{{./}}

\usepackage[
style=numeric-comp,
citestyle=numeric,
backend=biber,
firstinits=true,useprefix=true,sorting=none, sortcites=true,
bibencoding=utf-8,
maxnames=15,
minnames=1,
maxbibnames=15,minbibnames=14,
maxcitenames=2,mincitenames=2,
hyperref=auto,
backref=false,
backrefstyle=all+,
isbn=false,
eprint=false,
url=false,
doi=false]{biblatex}

 \renewbibmacro*{volume+number+eid}{\printfield{volume}\printfield{number}\printfield{eid}}

\DeclareNameAlias{sortname}{last-first}
\DeclareNameAlias{default}{last-first}

\renewbibmacro{in:}{}

\DeclareFieldFormat[article]{volume}{\mkbibbold{#1}}\DeclareFieldFormat[article]{number}{#1}\DeclareFieldFormat[article]{title}{#1} \DeclareFieldFormat[article]{pages}{#1}             \DeclareFieldFormat[article]{year}{(#1)}

\renewbibmacro*{journal+issuetitle}{\setunit{\addcomma\space \addspace}\printfield{journaltitle} \newunit}

\DeclareBibliographyDriver{article}{\usebibmacro{bibindex}\usebibmacro{begentry}\usebibmacro{author/translator+others}\setunit{\addcomma\space }\newblock \setunit{\addcomma\space} \printfield{title}
\newunit{\adddot\space } \printfield{journaltitle}
\setunit{}\newblock  
  \printfield{year}
  \setunit{\addspace} \iffieldundef{volume}{
  }
  {\printfield{volume}
  \setunit{\addcolon}}
  \printfield{number}
  \setunit{\addcolon}
  \printfield{pages}
  \newunit
\usebibmacro{finentry}}

\renewbibmacro*{book+year}{\setunit*{\space}\iffieldundef{book}
      {\usebibmacro{date}}
      {\printfield{issue}\setunit*{\addspace}\usebibmacro{date}}\newunit}

\DeclareFieldFormat[book]{title}{\mkbibemph{#1}}  \DeclareListFormat[book]{publisher}{#1}  \DeclareFieldFormat[book]{series}{\mkbibemph{#1}}  

\DeclareBibliographyDriver{book}{\usebibmacro{bibindex}\usebibmacro{begentry}\newunit\newblock
\printnames{author}
  \setunit{\adddot\space \addcomma\space } \usebibmacro{byauthor}\newunit
  \setunit{\adddot\space } \usebibmacro{title}\iffieldundef{series}
  {\skipentry}
  {\printfield{series}}
\newunit
  \setunit{\adddot\space } \printlist{publisher} \usebibmacro*{book+year}
  \usebibmacro{finentry}
}
  
\usepackage{color}
\usepackage{xcolor}
\usepackage{transparent} 

\definecolor{dark_blue}{RGB}{0,9,129}
\definecolor{dark_green}{RGB}{18,172,88} 
\definecolor{dark_blue_cite}{RGB}{0,84,168}
\definecolor{light_grey}{RGB}{190,190,190}

\definecolor{RedDark}{RGB}{139,0,0}
\definecolor{Red}{RGB}{255,0,0}
\definecolor{GreenDark}{RGB}{34,139,34} \definecolor{Green}{RGB}{50,205,50}

\definecolor{Blueone}{RGB}{0,96,173} \definecolor{Redone}{RGB}{173,77,0}

\RequirePackage{physics}
\newcommand{\vari}[1]{\widetilde{#1}}

\newcommand{\volumefraction}[0]{\ensuremath{\alpha}}
\newcommand{\volumefractionvar}[0]{\ensuremath{\vari{\alpha}}}
\newcommand{\massfraction}[0]{\ensuremath{Y}}
\newcommand{\massfractionvar}[0]{\ensuremath{\vari{Y}}}
\newcommand{\bu}[0]{\ensuremath{\vb{u}}}
\newcommand{\buvar}[0]{\ensuremath{\vari{\vb{u}}}}
\newcommand{\rhovar}[0]{\ensuremath{\vari{\rho}}}
\newcommand{\pseudomass}[0]{\ensuremath{m}}

\newcommand{\bx}[0]{\ensuremath{\vb{x}}}
\newcommand{\interfacialarea}[0]{\ensuremath{\Sigma}}
\newcommand{\interfacialareavar}[0]{\ensuremath{\vari{\Sigma}}}
\newcommand{\meancurvature}[0]{\ensuremath{H}}

\newcommand{\perturbation}[0]{\ensuremath{h}}
\newcommand{\perturbationvar}[0]{\ensuremath{\vari{h}}}
\newcommand{\pulsation}[0]{\ensuremath{w}}
\newcommand{\perturbmass}[0]{\ensuremath{m}}
\newcommand{\NJpot}[0]{\ensuremath{f}}
\newcommand{\kineticenergy}[0]{\ensuremath{\mathscr{K}}}
\newcommand{\potentialenergy}[0]{\ensuremath{\mathscr{U}}}

\newcommand{\Lagrangian}[0]{\ensuremath{L}}

\newcommand{\matdv}[1]{\ensuremath{D_t{#1}}}
\newcommand{\nbR}[0]{\ensuremath{\mathbb{R}}}

\newcommand{\Idmat}{\mathbb{I}_{d}}

\newcommand{\dinfeul}{\delta}  

 \newcommand{\action}{\mathscr{A}} 

\newcommand{\xlag}{\bs{X}} \newcommand{\xeul}{\bs{x}} 

\newcommand{\pathxeul}{\vb*{\varphi}} \newcommand{\pathxlag}{{\vb*{\varphi}^L}} 

\newcommand{\pathxlagvar}{\vb*{\varphi}^L}

\newcommand{\LAPvar}{\lambda}

\newcommand{\spaces}[1]{\IfEqCase{#1}{{M}{\mathcal{M}}{TM}{\mathcal{TM}}{TqM}{\mathcal{T}_{\pathM}\mathcal{M}}{TR}{\mathcal{TR}}{TR7}{\mathcal{TR}^{7}}{R7}{\mathbb{R}^{7}}{R3}{\mathbb{R}^{3}}{R}{\mathbb{R}}{Rplus}{\mathbb{R}^{+}}}[\PackageError{space}{Undefined option to space: #1}{}]}

\newcommand{\spacetx}[1]{\IfEqCase{#1}{{0}{\mathcal{V}}{t}{\mathcal{V}_{t}}{t0}{\mathcal{V}_{t_{0}}}{t1}{\mathcal{V}_{t_{1}}}}[\PackageError{spacetx}{Undefined option to spacetx: #1}{}]}

\newcommand{\fluidvolume}[0]{\ensuremath{\mathscr{B}}}

\newcommand{\beul}{b} 
\newcommand{\blag}{{b}^\text{L}}
\newcommand{\xtdomain}[0]{\Omega}

\DeclareCiteCommand{\citejournal}
  {\usebibmacro{prenote}}
  {\usebibmacro{citeindex}\usebibmacro{journal}}
  {\multicitedelim}
  {\usebibmacro{postnote}}

\newcommand{\citeay}[1]{\autocite{#1}} 
\usepackage[pdftex,breaklinks, urlcolor=blue,pdftitle={Work Report},pdfauthor={Pierre CORDESSE},pdfsubject={LaTeX}]{hyperref}
\hypersetup{
  colorlinks   = true,
  citecolor    = blue,
  linkcolor = blue,
}

\title{Derivation of a two-phase flow model with two-scale kinematics, geometric variables and surface tension using variational calculus}

\shorttitle{Two-phase flow modeling by means of variational calculus}

\author{P. Cordesse,
    \and
    S. Kokh,
    \and
    R. Di Battista,
    \and
    F. Drui,
    \and
    M. Massot
}

\shortauthor{P. Cordesse et~al.}

\begin{document}

\setcounter{page}{1}

\maketitle

\begin{abstract}
The present paper proposes a two-phase flow model that is able to account for two-scale kinematics and two-scale surface tension effects based on geometric variables at small scale. At large scale, the flow and the full geometry of the interface may be retrieved thanks to the bulk variables, while at small scale the interface  is accurately described by volume fraction, interfacial area density and mean curvature, called the geometric variables. Our work mainly relies on the Least Action Principle. The resulting system is an extension of a previous work modeling small scale pulsation in which surface tension was not taken into account at large or small scale. Whereas the original derivation assumes a cloud of monodispersed spherical bubbles, the present context allows for polydispersed, non-spherical bubbles. The resulting system of equations solely involves small scale geometric variables, thus contributing in the construction of a unified model describing both large and small scales.
\end{abstract}

\section{Introduction}\label{sec:introduction}
We consider, in this work, the simulation of two-phase flows involving potentially very different interface topologies in various areas of the flow. Such situations may occur in various industrial contexts including multiple stages, such as during the atomization of a liquid jet ranging from the deformation of the interface at the mouth of the injector, to the formation of a polydisperse spray of droplets downstream, through a mixed zone where the geometry of the interface may be complex.

The first step to pave the way to an unified Eulerian modeling of such complex flows was initiated 
in \cite{Drui_JFM_2019} by making a first connection between models for separate phase flows and subscale modeling, aiming at also describing disperse flows such a bubbly flows.
The starting ideas were to design a model that was able to account for two-scale kinematics and to close the model and identify small parameters in the case of bubbly flows, even if the final model was valid without any assumption on the structure of the subscale geometry of the interface.
Our goal here is two-fold: first, we propose to include more geometrical information on the subscale description than in \cite{Drui_JFM_2019}, relying on the fact that this information allows the description of a cloud of droplets in \cite{Essadki_2016,Essadki_Drui_2017} and 
second, we aim not only to account for two-scale kinematics but also for two-scale surface tension effects at both large scale and subscale.

This paper is structured as follows. First we present a careful choice of the variables that shall describe the evolution of the system. More precisely, an argument similar to the Tube Formula of Weyl~\cite{Weyl_1939,Gray_2004} will allow us to consider gas inclusions whose shape is diffeomorph to a sphere. This analysis will also allow us to draw connections between the density of interfacial area, the mean curvature and the volume fraction that will be considered as parameters of the flow. We will then design a Lagrangian energy that accounts for two-scale effects for both kinematics and capillarity effects. Finally we will use the Least Action Principle~\cite{Serrin_1959,Bedford_1978,Bedford_1985,Gavrilyuk_Gouin_1997,Gavrilyuk_1999} that will provide us with a set of governing equations for our two-phase medium.

\section{Description of the two-phase medium: two-scale structure}\label{sec:model}

\begin{figure}[h]
    \centering
      \begin{overpic}[width=0.8\textwidth ]{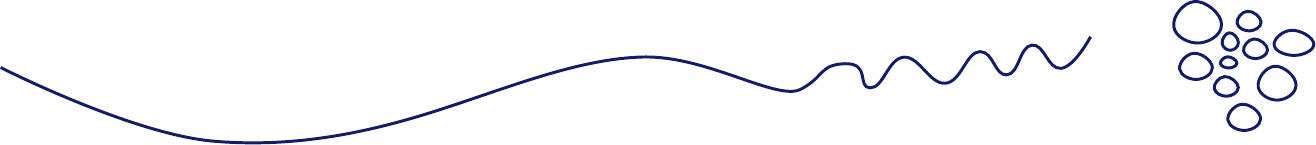} \put(20,-5){$\textcolor{dark_blue_cite}{\text{large scale}}$}
         \put(55,-5){$\textcolor{dark_blue_cite}{\text{large \& small scales}}$}
         \put(90,-5){$\textcolor{dark_blue_cite}{\text{small scale}}$}
      \end{overpic}
    \vspace{2em}
    \caption{Interface dynamics at large and small scales}
    \label{fig:interface_dynamics}
\end{figure}

We consider two compressible materials $k=1,2$ that are both equipped with a barotropic equation of state (EOS) $\rho_k \mapsto f_k$, where $f_k$ and $\rho_k$ are respectively the specific free energy and the density of the fluid. The partial pressure $p_k$ and the sound velocity $c_k$ of the component $k$ are defined by 
\begin{align*}
  p_k = \rho_k^2\dv*{f_k}{\rho_k}, && c_k^2 = \dv*{p_k}{\rho_k}.
\end{align*}
We assume that there is a velocity equilibrium between both components and we note $\bu$ as their common velocity. 
We assume our two-phase medium to be an immiscible mixture where, for each component, $Y_k$ and $\alpha_k$ denote the mass and volume fractions respectively. If $\rho$ is the density of the medium, we have
\begin{align*}
\rho = \volumefraction \rho_1 + (1-\volumefraction)\rho_2
,&&
\rho \massfraction_k = \volumefraction_k \rho_k,
\end{align*}
\vspace{-2em}
\begin{align*}
\text{with } \volumefraction = \volumefraction_1 = 1-\volumefraction_2 
,\ \massfraction = \massfraction_1 = 1-\massfraction_2.
\end{align*}
We now need to describe the properties of the interface that separates both materials and distribute these properties over two different scales. let us first underline that we assume both scales to be always simultaneously present at each time-position $(t,\xeul)$. In our approach, we assume that the preponderant scale dominates the interface effects in the model (see Figure~\ref{fig:interface_dynamics}). Let us now detail the interface assumption we shall use for both scales.

Concerning the large scale, we adopt a classic interface capturing approach (see for example \cite{Chanteperdrix_2002}): we suppose that the interface  position is captured within a narrow region where $\volumefraction$ rapidly varies from $\volumefraction\simeq 0$ to $\volumefraction\simeq 1$. When the large scale is dominant, we postulate that the outward unit normal to the interface is accurately given by $\grad\volumefraction/|\grad{} \volumefraction|$ as originally done in \cite{Brackbill_1992}.

When the small scale is predominant, we acknowledge that the properties of the interface are only available through the fields 
$(t,\bx)\mapsto\interfacialarea $ and 
$(t,\bx)\mapsto\meancurvature $
that respectively provide a measure of the interfacial density area and the mean curvature in the vicinity of $\bx$ at instant $t$. Let us now specify
how these fields are related to the other variables of the flow. First, we assume that the topology of the small scale is consistent with a population of small gas inclusions whose shape is diffeomorph to a sphere in such way that we can apply Weyl's Tube Formula \cite{Weyl_1939,Gray_2004}.

We now suppose that the shape of the bubbles whose position is $\bx$ at instant $t$ is altered by a small scale perturbation that is normal to their surface and with a (signed) length magnitude (oriented with respect to the outward unit normal) $h(t,\bx)$ as depicted in Figure~\ref{fig:closed_mapping}. 
In other words, $(t,\bx)\mapsto h$ can be interpreted as a first-order estimate of the deformation length for the bubbles located in the vicinity of $\bx$ at instant $t$.
For such perturbations, the normal variation of the 
surface and the volume of the inclusions can be connected to the mean curvature and the surface of the inclusions as detailed in Appendix~\ref{sec: tube formula}. Consequently we postulate that the fields $\volumefraction$, $\interfacialarea$, $\meancurvature$, $h$ are connected through the following relations:
\begin{align}\label{eq:tube_perturbation}
\dv{\interfacialarea}{\perturbation} &= - 2 {\meancurvature} \interfacialarea ,&
\dv{\volumefraction}{\perturbation} &= \interfacialarea
.
\end{align}
Hypotheses~\eqref{eq:tube_perturbation} express the fact that a reminiscence of relation~\eqref{eq: formule tube like relation} is valid for each small-scale gas inclusion through the fields $\volumefraction$, $\interfacialarea$, $\meancurvature$, $h$. In the sequel we make the strong assumption that $(t, \xeul) \mapsto \meancurvature$ is given a priori and is not altered by the flow. 
This assumption can be lifted, but requires a more involved modeling at the subscale level. 

We conclude this section by stating the following constraints on the flow
\begin{subequations}\label{eq: hyp for rho Y conservation and alpha sigma}
\begin{align}
\pdv{\rho}{t} + \div{(\rho\bu)} &=0 
,& 
\pdv{\rho \massfraction}{t} + \div{(\rho \massfraction \bu)} &=0,
\\
 \matdv{\interfacialarea} + 2 \meancurvature \interfacialarea \matdv{\perturbation}
 &= 0
 , & 
 \matdv{\volumefraction} - \interfacialarea \matdv{\perturbation}
 &=0
\end{align}
\end{subequations}
that respectively pertain to total mass conservation, partial mass conservation and the constrained evolution of $\volumefraction$, $\interfacialarea$ and $h$ through the topological requirement~\eqref{eq:tube_perturbation}.
\begin{remark} The small scale perturbation encompasses purely normal variation of the inclusion interface. One could also account for volume stretching generated by a tangential variation of the normal as proposed in \cite{Lhuillier_1985}. However this situation is out of the scope of the present study since it requires an additional potential fluctuation of velocity at the subscale level, which we have not taken into account.
\end{remark}
\section{Two-scale Lagrangian energy}
In order to derive a system of equations for our two-phase flow we exploit the Least Action Principle following \cite{Serrin_1959,Bedford_1978,Bedford_1985,Gavrilyuk_Gouin_1997,Gavrilyuk_1999}. The first step consists in providing the medium with a Lagrangian energy $\Lagrangian$. Following classic lines, we define $\Lagrangian$ to be the difference between the kinetic energy and the potential energy of the system. In order to account for both large and small scale phenomena we will suppose that the kinetic energy of the system is defined by $ \kineticenergy^\text{small} + \kineticenergy^\text{large} $
where $\kineticenergy^\text{small}$ and  $\kineticenergy^\text{large}$ 
refer respectively to the small scale and the large scale kinetic energy. In the same way, we assume that the potential energy of the system is 
$ 
\potentialenergy{}_\text{bulk}
+ 
\potentialenergy{}_\text{int}^\text{large}
+ 
\potentialenergy{}_\text{int}^\text{small}
$ 
where
$\potentialenergy{}_\text{bulk}$, denotes the bulk potential 
energy of the system, 
$\potentialenergy{}_\text{int}^\text{large}$ and $\potentialenergy{}_\text{int}^\text{small}$ are interfacial energies associated with large and small scale interface descriptions respectively.

The large scale kinetic energy is defined by setting
\begin{equation}
\kineticenergy^\text{large} = \frac{1}{2} \rho |\bu|^2
.
\label{eq: def large scale kinetic energy}
\end{equation}
We leverage the small scale variations of the inclusions $h$ in order to define $\kineticenergy^\text{small}$ as follows
\begin{align}\label{eq:energy_kin_small}
  \kineticenergy^\text{small} = \frac{1}{2} 
  \pseudomass(\volumefraction, \interfacialarea) 
  \vert \matdv{\perturbation} \vert^{2}
\end{align}
where $\matdv\cdot = \partial_t \cdot + \bu^t\grad{}\cdot$ and
$m$ has the same dimensions as a density. The energy contribution (\ref{eq:energy_kin_small}) can be thought to be related to the virtual mass energy associated to the volume deformation of a gas inclusion at small scale. For the sake of simplicity we shall assume in the following sections that $\pseudomass$ is a constant.

We now turn to the definition of the potential energy of the system. We make the standard assumption that the bulk potential energy can be expressed as a bulk free energy as follows
\begin{equation}
\potentialenergy_\text{bulk} = \rho \NJpot(\rho,Y,\volumefraction).
\end{equation}
For the large scale interface energy we postulate that 
\begin{equation}\label{eq:energy_pot_large_interface}
  \potentialenergy_\text{int}^\text{large}
   = \frac{1}{2} 
   \sigma \vert \grad{\volumefraction} \vert^{2}
   ,
\end{equation}
where $\sigma >0 $ is a coefficient pertaining to the large-scale capillarity. We define the interface energy associated with the small scale by
\begin{align}\label{eq:energy_pot_small_interface}
  \potentialenergy_\text{int}^\text{small} = \gamma \interfacialarea,
\end{align}
where $\gamma>0$ is a coefficient related to small scale capillarity. For the sake of simplicity we shall assume in the sequel that both $\sigma$ and $\gamma$ are constant.

Finally, the resulting Lagrangian $\Lagrangian$ can be expressed as a function 
of
$\rho$, $\massfraction$, $\bu$, $\matdv{\perturbation}$, 
$\volumefraction$, 
$\interfacialarea$ and
$\grad{\volumefraction}$
as follows
\begin{equation}
\label{eq:lagrangian_energy}
\Lagrangian(\rho, \massfraction, \bu, \matdv{\perturbation}, \volumefraction, \interfacialarea, \grad{\volumefraction})
= 
\frac{\rho }{2}  \vert \bu \vert^{2}
+ 
\frac{\pseudomass(\volumefraction, \interfacialarea) }{2} 
\vert \matdv{\perturbation} \vert^{2}  
- \rho \NJpot(\rho, \massfraction, \volumefraction)
- \frac{\sigma }{2} \vert \grad{\volumefraction} \vert^{2} 
-\gamma\interfacialarea
.
\end{equation}

\begin{remark}
It is important to emphasize that we do not provide here any mechanism for distributing the information carried by the variables of the system between
small and large scales, which is crucial but out of the scope of the present work.
\end{remark}\section{Extremization of the Action}
We now follow classic lines of the Least Action Principle. Consider $\fluidvolume(t)\subset\nbR{}^3$ the volume occupied by the fluid for $t\in[t_0,t_1]$. Let $\xlag\in\fluidvolume(t_0) $ be the Lagrangian coordinates associated with the reference frame at instant $t=t_0$, then we note $(t,\xlag)\mapsto\pathxlag$ the position of the fluid particle whose position is $\xlag$ at $t=t_0$. If $(t,\xeul)\mapsto \beul$ is any Eulerian field, it can be associated with the Lagrangian field $(t,\xlag)\mapsto \blag$ by setting 
$\beul(\pathxlag(\xlag,t),t) = \blag(\xlag,t) $.
As $h$ can be deduced from $\volumefraction$ and $\interfacialarea$ by 
\eqref{eq: hyp for rho Y conservation and alpha sigma}, then the flow can be fully characterized by 
$(t,\xeul)\mapsto (\rho,\bu,\massfraction,\volumefraction,\interfacialarea)$ 
or equivalently by 
$\xeul\mapsto(\massfraction,\volumefraction,\interfacialarea)$ and
$(t,\xlag)\mapsto \pathxlag$ if 
$\pathxlag$ complies with the mass conservation equation. 

For a given transformation of the medium 
$\xeul\mapsto(\massfraction,\volumefraction,\interfacialarea)$ and
$(t,\xlag)\mapsto \pathxlag$
 let
$(t,\xeul,\LAPvar)\mapsto(\massfractionvar,\volumefractionvar,\interfacialareavar)$ and
$(t,\xlag,\LAPvar)\mapsto \pathxlagvar$ be a family of medium transformations 
parametrized by $\LAPvar\in[0,1]$. We note
$ \vari{\xtdomain}(\LAPvar)=\big\{
    (t,\vari{\pathxlagvar}(t,\xlag,\LAPvar))|\xlag\in\fluidvolume(t_0), t\in[t_0,t_1]
    \big\}
$ and we require these fields to satisfy constraints pertaining to mass conservation
\begin{subequations}
\label{eq: family transformation constraints}
\begin{align}
\pdv{\rhovar}{t} + \div{(\rhovar \buvar)} &=0 
,&
\pdv{\rhovar \massfractionvar}{t} + \div{(\rhovar \massfractionvar \buvar)} 
&=0 
,
\intertext{supplemented by constraints regarding the topology evolution}
\matdv{\interfacialareavar} 
&=
  - 2 \meancurvature \, \interfacialareavar \, \matdv{\perturbationvar} 
,&
\matdv{\volumefractionvar} &=
  \interfacialareavar \, \matdv{\perturbationvar},
\end{align}

\end{subequations}and finally classic boundary constraints
\begin{align}
  (\massfractionvar,\volumefractionvar,\interfacialareavar)(t,\xeul,\LAPvar=0,1)
 &=
  (\massfraction,\volumefraction,\interfacialarea)(t,\xeul)
  ,&
  \vari{\pathxlagvar}(\xlag,t,\LAPvar=0,1) &=   \pathxlag(\xlag,t),
\end{align}  
\begin{align}
(\massfractionvar,\volumefractionvar,\interfacialareavar)(t,\xeul,\LAPvar)
 &=g
  (\massfraction,\volumefraction,\interfacialarea)(t,\xeul)
    ,&
    (t,\xeul)&\in\partial \vari{\xtdomain}(\LAPvar),
  \\
  \vari{\pathxlagvar}(\xlag,t,\LAPvar) &=   \pathxlag(\xlag,t)
  ,&
  (t,\xlag)&
  \in\partial \qty([t_0,t_1]\times \fluidvolume(t_0)).
\end{align}
Following standard lines, this family of transformation yields a family of infinitesimal transformations defined as follows
\begin{align}
\dinfeul_{\LAPvar}\pathxeul(t,\pathxlag(t,\xlag))
&=
\qty(\pdv{\pathxlagvar}{\LAPvar})_{\xlag,t}(t,\xlag,\LAPvar=0)
,&
\dinfeul_{\LAPvar} b(t,\xeul)
&=
\qty(\pdv{\vari{b}}{\LAPvar})_{t,\xeul}(t,\xeul,\LAPvar=0)
,
\label{eq: inifinitesimal variations definition}
\end{align}
for $b\in\{\rho,\massfraction,\volumefraction,\interfacialarea,\bu\}$.
Let us now define the Hamiltonian action $\action$ associated with $\xtdomain$ for the family of transformations 
$(t,\xeul,\LAPvar)\mapsto(\massfractionvar,\volumefractionvar,\interfacialareavar)$ and
$(t,\xlag,\LAPvar)\mapsto \vari{\pathxlagvar}$ 
\begin{equation}
\label{eq: action definition}
\action(\LAPvar) = \int_{\vari{\xtdomain}(\LAPvar)}
\Lagrangian(\vari{\rho}, \massfractionvar, \buvar, \matdv{\perturbationvar}, \volumefractionvar, \interfacialareavar, \grad{\volumefractionvar})\,
\dd\xeul\dd t
.
\end{equation}

The Least Action Principle states that a physical transformation of the system verifies
\begin{equation}
\label{eq: action is extremal}
\dv{\action}{\LAPvar} (0) = 0.
\end{equation}
Relation~\eqref{eq: action is extremal} will provide the motion equations of the flow. In order to obtain a set of partial differential equations, we need to express $\dv*{\action}{\LAPvar}$. 
Using definition~\eqref{eq: inifinitesimal variations definition} we can write
\begin{multline}
\label{eq:first_var_action}
  \dv{\action}{\LAPvar} (0) 
  = \int_{\Omega(0)} 
  \left[
    \pdv{\Lagrangian}{\rho} \dinfeul_{\LAPvar} {\rho} + \pdv{\Lagrangian}{\massfraction} 
  \dinfeul_{\LAPvar} {\massfraction} +  \pdv{\Lagrangian}{\bu} \dinfeul_{\LAPvar} {\bu} + \pdv{\Lagrangian}{(\matdv{\perturbation})} \dinfeul_{\LAPvar}({\matdv{\perturbation}}) \right.
  \\
  \left.
  + \pdv{\Lagrangian}{\volumefraction} 
  \dinfeul_{\LAPvar} {\volumefraction} +  \pdv{\Lagrangian}{\interfacialarea} 
  \dinfeul_{\LAPvar} {\interfacialarea} + \pdv{\Lagrangian}{(\grad{\volumefraction})} 
  \dinfeul_{\LAPvar} ({\grad{\volumefraction}}) \right]
  \, \dd\xeul \dd t 
  .
\end{multline}
Applying \eqref{eq: inifinitesimal variations definition}
with the constraints~\eqref{eq: family transformation constraints} allows to express following relations between the infinitesimal variations 
\begin{subequations}
\begin{align}
\label{eq:inf_var_density}
  \dinfeul_{\LAPvar} \rho &= - \div{\left(\rho \dinfeul_{\LAPvar} \pathxeul \right)}
,\\
\label{eq:inf_var_massfrac}
  \dinfeul_{\LAPvar} Y &= - \grad{Y}^{T} \dinfeul_{\LAPvar} \pathxeul
,\\
\label{eq:inf_var_velocity}
  \dinfeul_{\LAPvar} \bu &=
   \matdv{\left(
   \dinfeul_{\LAPvar} \pathxeul \right)} 
   - \left( \dinfeul_{\LAPvar} \pathxeul^{T} \grad{} \right) \bu 
,\\
\label{eq:inf_var_perturbation}
  \dinfeul_{\LAPvar} \left(  \matdv{\perturbation} \right) &= \frac{1}{\interfacialarea} \dinfeul_{\LAPvar} \left( \matdv{ \volumefraction}  \right) - \frac{\matdv{ \perturbation }}{\interfacialarea} \dinfeul_{\LAPvar} \interfacialarea
.
\end{align}
\label{eq: infinitesimal variations relations}
\end{subequations}Recasting relations~\eqref{eq: infinitesimal variations relations} into \eqref{eq:first_var_action}
provides
\begin{align}
   &\int_{\Omega(0)} [
  \textbf{A}^{T} \dinfeul{\pathxeul} + \text{B} \, \dinfeul{\volumefraction} + \text{C} \, \dinfeul{\interfacialarea}
    ]\,\dd\xeul\dd t
    =0,
\end{align}
\begin{align}
\begin{IEEEeqnarraybox}[\IEEEeqnarraystrutmode
\IEEEeqnarraystrutsizeadd{1pt}{1pt}][c]{lc}
\textbf{A}^{T}  &= \partial_{t} \left( \frac{\partial \Lagrangian }{\partial \bu } \right)
+ \div{\left[ \right(\frac{\partial \Lagrangian }{\partial \bu}\left)^{T} \, \bu^{T} \right]}
+ (\grad{\bu} )^{T} \left(\frac{\partial \Lagrangian }{\partial \bu }\right)^{T}
+ \partial_{t}\left(  \frac{1}{\interfacialarea} \frac{\partial \Lagrangian }{\partial \matdv{\perturbation}} \grad{\alpha} \right) 
\\
&+ \div{\left[  \frac{1}{\interfacialarea} \frac{\partial \Lagrangian }{\partial \matdv{\perturbation}} \grad{\alpha}  \, \bu \right]}
+ \frac{1}{\interfacialarea} \frac{\partial \Lagrangian }{\partial \matdv{\perturbation}}  (\grad{\bu})^{T}  \grad{\volumefraction}
- \rho \left(\grad{ \left[ \frac{\partial \Lagrangian }{\partial \rho} \right]}\right)^{T}
+ \frac{\partial \Lagrangian }{\partial \massfraction} \grad{\massfraction}, 
\end{IEEEeqnarraybox}
\end{align}
\begin{align}
\begin{IEEEeqnarraybox}[\IEEEeqnarraystrutmode
\IEEEeqnarraystrutsizeadd{1pt}{1pt}][c]{cl}
\text{B}  &= \frac{\partial \Lagrangian }{\partial \volumefraction} - \div{ \left[ \frac{\partial \Lagrangian}{\partial \grad{\volumefraction} } \right] } - \partial_{t}\left(  \frac{1}{\interfacialarea} \frac{\partial \Lagrangian }{\partial \matdv{\perturbation}} \right) - \div{\left[ \frac{1}{\interfacialarea} \frac{\partial \Lagrangian }{\partial \matdv{\perturbation}}\, \bu \right]},
\end{IEEEeqnarraybox}
\end{align}
\begin{align}
\begin{IEEEeqnarraybox}[\IEEEeqnarraystrutmode
\IEEEeqnarraystrutsizeadd{1pt}{1pt}][c]{cl}
\text{C}  &= \frac{\partial \Lagrangian }{\partial \interfacialarea} -  \frac{1}{\interfacialarea} \frac{\partial \Lagrangian }{\partial \matdv{\perturbation}} \matdv{\perturbation}.
\end{IEEEeqnarraybox}
\end{align}
We can conclude that the Least Action Principles applied to the Lagrangian energy  defined by \eqref{eq:lagrangian_energy} yields the following equations of motion
\begin{subequations}\label{eq: motion eq first form}
\begin{align}
A=0, && B=0, && C=0.
\end{align}
\end{subequations}
Let us further express the equations of motions into a more familiar form. With the definition \eqref{eq:lagrangian_energy} of $\Lagrangian$ one then has
\begin{subequations}\label{eq:lagrangian_diff}
\begin{align}
\pdv{\Lagrangian}{\rho} 
&= 
\frac{\vert \bu \vert^{2}}{2} - \NJpot - \rho \pdv{\NJpot}{\rho} 
,&
\pdv{\Lagrangian}{\massfraction} 
&= 
- \rho \pdv{\NJpot}{\massfraction} 
,&
\pdv{\Lagrangian}{\bu} &= \rho \bu 
,\\
\pdv{\Lagrangian}{\left(\matdv{\perturbation}\right)} 
&= \pseudomass \matdv{\perturbation} 
,&
\pdv{\Lagrangian}{\interfacialarea} 
&= -\gamma 
,&
\pdv{\Lagrangian}{\volumefraction} 
&= 
- \rho \pdv{\NJpot}{\volumefraction} 
,&
\pdv{\Lagrangian}{\left( \grad{\volumefraction} \right)} 
&= \sigma \grad{\volumefraction}
.
\end{align}
\end{subequations}
We obtain 
\begin{align}
\begin{IEEEeqnarraybox}[\IEEEeqnarraystrutmode
\IEEEeqnarraystrutsizeadd{1pt}{1pt}][c]{lc}
\textbf{A}  &= \partial_{t} \left( \rho \bu \right)
+ \div{\left[ \rho \bu \bu^{T} \right]}
+ \rho \grad{\bu} \cdot \bu
+ \partial_{t}\left(  \frac{\pseudomass}{\interfacialarea} \matdv{\perturbation} \grad{\alpha}\right)
+ \div{\left[  \frac{\pseudomass}{\interfacialarea} \matdv{\perturbation}  \grad{\alpha} \, \bu \right]}
\\
& + \frac{\pseudomass}{\interfacialarea} \matdv{\perturbation} (\grad{\bu})^{T} \grad{\volumefraction} 
- \rho \grad{\left[ \frac{1}{2} \vert \bu \vert^{2} - \NJpot - \rho \partial_{\rho} \NJpot \right]} 
- \rho \partial_{\massfraction} \NJpot \grad{\massfraction}, 
\end{IEEEeqnarraybox}
\end{align}
\begin{align}
\begin{IEEEeqnarraybox}[\IEEEeqnarraystrutmode
\IEEEeqnarraystrutsizeadd{1pt}{1pt}][c]{c}
\text{B}  = \partial_{t}\left( \frac{\pseudomass}{\interfacialarea} \matdv{\perturbation} \right) + \div{\left[ \frac{\pseudomass}{\interfacialarea} \matdv{\perturbation} \, \bu \right]} + \rho \partial_{\volumefraction}\NJpot + \div{\left[\sigma  \grad{\volumefraction}\right]},
\end{IEEEeqnarraybox}
\end{align}
\begin{align}
\begin{IEEEeqnarraybox}[\IEEEeqnarraystrutmode
\IEEEeqnarraystrutsizeadd{1pt}{1pt}][c]{c}
\text{C}  = - \gamma - \frac{\pseudomass}{\interfacialarea}\left(\matdv{\perturbation}\right)^{2}.  
\end{IEEEeqnarraybox}
\end{align}
\section{Final form of the system}
We define the pressure $p$ of the two-phase medium and the partial pressures $p_{k}$ of each phase and introduce a new variable $\pulsation$ by setting 
\begin{align}
p &=  \rho^{2} \pdv{f}{\rho}
,&
p_{k} &=  \rho_{k}^{2} \pdv{f}{\rho_{k}}
,&
\pulsation &= \frac{\matdv{\volumefraction}}{\rho \massfraction \interfacialarea^2},
\end{align}
then by injecting relations~\eqref{eq:lagrangian_diff} into \eqref{eq: motion eq first form} one obtains the system
\begin{subequations}\label{eq:final_system}
\begin{empheq}[left=\empheqlbrace]{align}
\pdv{\rho}{t} + \div{ \left[ \rho \bu \right]} &= 0 
,\\
\pdv{\rho Y}{t} + \div{ \left[ \rho Y \bu  \right]} &=0
,\\
\pdv{\rho \bu}{t} + \div{\left[(\rho \bu \bu^{T}) + \left( p + \frac{1}{2} \frac{\pseudomass}{\interfacialarea^{2}} \left(\matdv{\volumefraction}\right)^{2} \right) \Idmat + \sigma \grad{\volumefraction}\grad{\volumefraction}^{T}- \sigma \frac{\vert \grad{\volumefraction} \vert^{2}}{2}\Idmat \right]}&=0 \label{eq:final_system_large_scale_mom}
,\\
\matdv{\volumefraction} - \rho \massfraction \pulsation \interfacialarea^{2}
&=0
,\\
\matdv{\pulsation} + \frac{1}{\rho \massfraction \perturbmass } \left( p_{2}-p_{1}  + \div{\left[\sigma \grad{\volumefraction}\right]  } \right)  &= 0 \label{eq:final_system_small_scale_mom}
,\\
\matdv{\left( \rho \interfacialarea\right)} 
-
 \frac{2\rho \massfraction \pulsation \interfacialarea }{\gamma} 
 \left( 
 p_{2}-p_{1} 
 +  \div{\left[ \sigma \grad{\volumefraction}\right]} 
 \right) &= 0
.
\end{empheq}
\end{subequations}
\begin{remark}
System~\eqref{eq:final_system} is a generalization of the system found in \cite{Drui_JFM_2019} and degenerates towards it when considering the interfacial area density as a function of the volume fraction only and using the notation of the author one identifies $\nu = \pseudomass/\interfacialarea$. However the variable $\pulsation$ is not the defined as the pulsation $w$ found in \cite{Drui_JFM_2019}
\end{remark}
\begin{remark}
System~\eqref{eq:final_system} is valid for any flow topology.
\end{remark}

In the large scale momentum equation~\eqref{eq:final_system_large_scale_mom}, the terms function of the volume fraction gradient are common terms found in the literature \citeay{Blanchard_2016}.

Equation~\eqref{eq:final_system_small_scale_mom} is a small scale momentum equation on the variable $\pulsation$ which can be interpreted as the small scale pulsation of any structures.

Neglecting second order terms and capillarity at large scale   ($\sigma=0$), the spectrum of System~\eqref{eq:final_system} yields $(u,u,u,u,u \pm \sqrt{\partial_{\rho} \Pi + \interfacialarea/\rho \partial_{\interfacialarea} \Pi})$ with $\Pi = p + 1/2 \pseudomass \rho^{2} \massfraction^{2}  \pulsation^{2} \interfacialarea^{2}$. The sound speed is impacted by the small scale effects.

\section{Conclusion}
In the present contribution, we have designed a two-phase flow model that is able to account for two-scale kinematics and two-scale surface tension effects as well as subscale pulsation momentum
using the Least Action Principle. The system obtained is an extension of the work of \cite{Drui_JFM_2019} and degenerates naturally towards the system proposed in it.

This is solid foundation on which to build a dissipative structure for the present model by using an entropy evolution equation as in \cite{Drui_JFM_2019} and relying also on the recent work \cite{Cordesse_2019_CMS}. We are also investigating the possibility of extending the degree at which the system is out of equilibrium, in particular as far as velocity of the phases are concerned \cite{Cordesse_2018_c}.
Including a more detailed description of the subscale geometry and topology in order to be consistent with \cite{Essadki_2016,Essadki_Drui_2017} and providing a clear mechanism for distributing the information carried by the variables of the system over small and large scales are still a key issue and the subject of our current research.

\subsection*{Acknowledgment}
The research of P.~Cordesse  and R.~Di~Battista are supported by a CNES/ONERA  and a DGA/Ecole polytechnique PhD grants respectively.
The support of IFPEn, Ecole polytechnique and Initiative HPC@Maths (PIs F.~Alouges and M.~Massot) for M. Massot and R. Di Battista, and of CEA for Samuel Kokh are gratefully acknowledged.  This work was conducted during the Summer Program 2018 at NASA Ames Research Center and the support and precious help of N.N.~Mansour and F.~Panerai are also gratefully acknowledged. 
\appendix

\newcommand{\surface}[0]{\mathcal{S}}
\newcommand{\domain}[0]{\mathcal{D}}
\newcommand{\surfacemapping}[0]{\mathcal{M}}
\newcommand{\interval}[0]{I}
\newcommand{\measure}[0]{\text{meas}}
\newcommand{\localcurvature}[0]{H^\text{loc}}
\newcommand{\normal}[0]{\vb*{n}}

\section{Normal perturbation of a regular closed surface}
\label{sec: tube formula}

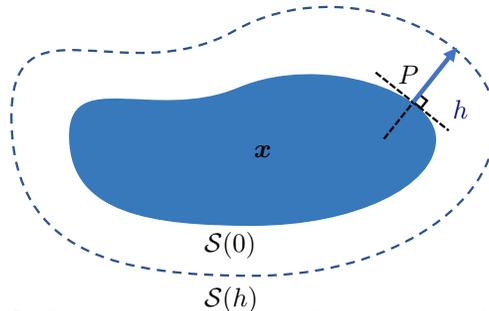
\begin{figure}[h]
    \centering
      \begin{overpic}[width=0.4\textwidth ]{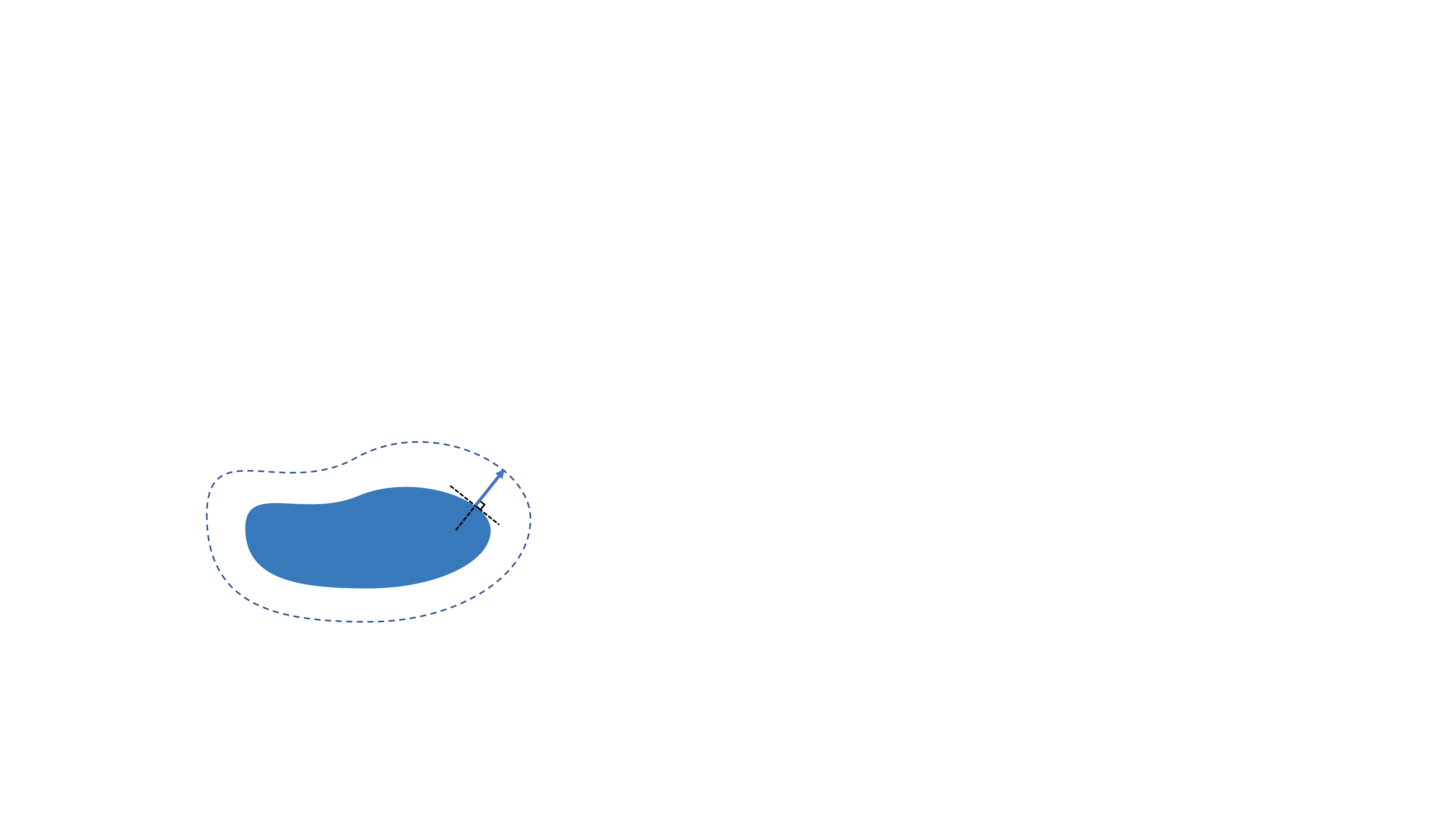} \put(40,6){$\surface(0)$}
         \put(40,-5){$\surface(h)$}
\put(79,40){$P$}
         \put(50,25){$\bs{x}$}
\put(90,33){$\textcolor{dark_blue}{\perturbation}$}
      \end{overpic}
    \caption{Closed volume undergoing normal variation}
    \label{fig:closed_mapping}
\end{figure}

Let $\domain$ be an open subset of \( \nbR{}^2 \) and \(\interval{}\) 
be an interval of \(\nbR{}\).
Consider a regular closed surface \(\surface \)
defined by the mapping 
\(
(u,v)\in\domain \mapsto
\surfacemapping(u,v) \in\nbR{}^3
\). We note 
\(  \normal(u,v)\in \nbR^3  \) the unit outward normal to \(\surface \) at 
the point \( \surfacemapping(u,v) \in\surface \).
Let us now consider
a family of surfaces 
\( 
\surface{}(h)
=
 \{ \surfacemapping(u,v) + h \, \normal(u,v)\in\nbR{}^3 ~|~ (u,v)\in \domain\} 
\)  parametrized by \(h\in\interval\) where \(\surfacemapping \) is a smooth mapping as depicted in Figure~\ref{fig:closed_mapping}. 
Following~\cite{Gray_2004}, one can show that
\begin{equation}
\measure[\surface(h)]
=
\measure[\surface(0)]
-
2 h
\int_{P\in\surface} \localcurvature(P) \dd P
+
O(h^2),
\label{eq: tube relation like}
\end{equation}
 where \(\dd P \) is the standard surface measure defined over \(\surface\) and \(\localcurvature(P) \) is the mean curvature of \(\surface\) at the point \(P\in\surface \). Let us define the average mean curvature \( \localcurvature_{\surface}\) of \( {\measure(\surface)}\) by
\[
\localcurvature_{\surface}  
=
\frac{\int_{P\in\surface} \localcurvature(P) \dd P}
{\measure[\surface]}
.
\]
From \eqref{eq: tube relation like} one deduces that
\begin{equation}
\dv{}{h}\qty(\measure[\surface(h)])
=
-2
\localcurvature_{\surface}  
{\measure[\surface]}
.
\label{eq: formule tube like relation}
\end{equation}

\printbibliography

\end{document}